\documentclass[aps, prl, amsmath, amssymb]{revtex4}

\begin{document}

\title{Operator Coupling of Gauge Fields and Unparticles}
\author{A. Lewis Licht}
\affiliation{Dept. of Physics\\U. of Illinois at Chicago\\Chicago, 
Illinois 60607\\licht@uic.edu}

\begin{abstract}
We show that it is possible to couple gauge fields to unparticles without the use of path 
integrals in the unparticle effective action.  This is done by 
treating the unparticle field as a vector in an abstract Hilbert 
space, and the gauge field as a linear operator on that space.
\end{abstract}

\maketitle

\section{Introduction}\label{S:intro}
Terning et al~\cite{Tern} have used a path integral in 
Georgi's~\cite{Geo-1}~\cite{Geo-2} unparticle effective action to 
couple gauge fields to the unparticles.  We have shown in a related 
work~\cite{Li} that if done rigorously, this leads to a formula for 
the gauge-unparticle vertex that is much more complicated than 
that found in Ref.~\cite{Tern}.  Here we show that there is an 
operator formulism, not involving path integrals, that does lead to 
Terning et al's result.  

\section{Operator powers as branch cut integrals}\label{S:bra}
The unparticle formulism relies very heavily on the representation of 
operator powers as branch cut integrals.  Given the complex function 
$f\left( z \right) = z^n $, we can express it as a contour integral,
\begin{equation}
z^n  = \frac{1}
{{2\pi i}}\oint {dz'\frac{{z'^n }}
{{z' - z}}} 
\end{equation}
If $ - 1 < n < 0$ we can expand the contour to wrap around the 
positive x-axis and we can discard the loop integral at infinity.  
Then
\begin{equation}
z^n  =  - \frac{{e^{\pi in} }}
{\pi }\sin \left( {\pi n} \right)\int_0^\infty  {dx\frac{{x^n }}
{{x - z}}} 
\end{equation}
If n is not in this range, say $n = k - \delta$, where k is an 
integer and $\delta$ is in the range, one could still use 
the branch integral to express $z^{-\delta}$, and just multiply by 
$z^{k}$.  In many cases it should be possible to start with n in the 
negative unit range, do the branch cut integral, and analytically 
continue to the desired value of n.

Georgi~\cite{Geo-1}~\cite{Geo-2} has for the propagator of his 
unparticle field;
\begin{equation}
\begin{gathered}
  \left\langle {0\left| {T\left( {\Phi _u^\dag  \left( x \right)\Phi _u \left( 0 \right)} \right)} \right|0} \right\rangle  = S\left( {x,0} \right) \\ 
   = \int {\frac{{d^4 p}}
{{\left( {2\pi } \right)^4 }}e^{ - ipx} \frac{{A_{du} }}
{{2\pi i}}\int_0^\infty  {dM^2 \frac{{\left( {M^2 } \right)^{d_u  - 2} }}
{{M^2  - p^2  - i\varepsilon }}} }  \\ 
   = \int {\frac{{d^4 p}}
{{\left( {2\pi } \right)^4 }}e^{ - ipx} \frac{{A_{du} e^{ - \pi id_u } }}
{{ - 2i\sin \left( {\pi d_u } \right)}}} \left( {p^2  + i\varepsilon } \right)^{d_u  - 2}  \\ 
\end{gathered} 
\end{equation}
Where
\begin{equation}
A_{du}  = \frac{{16\pi ^{5/2} }}
{{\left( {2\pi } \right)^{2d_u } }}\frac{{\Gamma \left( {d_u  + 1/2} \right)}}
{{\Gamma \left( {d_u  - 1} \right)\Gamma \left( {2d_u } \right)}}
\end{equation}
The quadratic part of the unparticle action integral is then nonlocal,
\begin{equation}
I  = \int {d^4 xd^4 y\Phi _u^\dag  \left( x \right)K\left( {x,y} \right)} \Phi _u (y)
\end{equation}
Where K and S are inverse operators,
\begin{equation}
\int {d^4 zK\left( {x,z} \right)S\left( {z,y} \right)}  = i\delta ^4 \left( {x - y} \right)
\end{equation}
The Kernel K can be expressed as:
\begin{equation}
K\left( {x,y} \right) = \frac{{2\sin \left( {\pi d_u } \right)e^{\pi id_u } }}
{{A_{du} }}\int {\frac{{d^4 p}}
{{\left( {2\pi } \right)^4 }}e^{ - ip\left( {x - y} \right)} } \left( {p^2  + i\varepsilon } \right)^{2 - d_u } 
\end{equation}
We can express K in terms of a branch cut integral
\begin{equation}
K\left( {x,y} \right) = \frac{{2\sin ^2 \left( {\pi d_u } \right)}}
{{\pi A_{du} }}\int {\frac{{d^4 p}}
{{\left( {2\pi } \right)^4 }}} e^{ - ip\left( {x - y} \right)} \int_0^\infty  {dM^2 \frac{{\left( {M^2 } \right)^{2 - d_u } }}
{{M^2  - p^2  - i\varepsilon }}} 
\end{equation}

Where we have assumed an analytical continuation to the appropriate value of 
$d_{u}$. 

\section{The Operator Formulism}\label{S:OpF}
We wish to have the action invariant under the gauge transformation:
\begin{equation}
\begin{gathered}
  \Phi _u \left( x \right) \to e^{ig\Lambda \left( x \right)} \Phi _u \left( x \right) \hfill \\
  A_\mu  \left( x \right) \to A_\mu  \left( x \right) + \partial _\mu  \Lambda \left( x \right) \hfill \\ 
\end{gathered} 
\end{equation}
We do this by writing the action integral as an inner product in a 
Hilbert space, and formulating derivatives and functional 
multiplication as operators on that Hilbert space.  

We consider the functions $\Phi _u \left( x \right)$ as corresponding 
to Hilbert space vectors,
\begin{equation}
\Phi _u \left( x \right) = \left\langle {x\left| {\Phi _u } \right.} \right\rangle 
\end{equation}
where the $\vert x >$ are eigenkets of a position operator:
\begin{equation}
\begin{gathered}
  X^\mu  \left| x \right\rangle  = x^\mu  \left| x \right\rangle  \\ 
  \left\langle x \right|\left. {x'} \right\rangle  = \delta ^4 \left( {x - x'} \right) \\ 
\end{gathered} 
\end{equation}
We define a momentum operator $P_{\nu}$ so that
\begin{equation}
\left[ {X^\mu  ,P_\nu  } \right] =  - i\delta _\nu ^\mu  
\end{equation}
The eigenkets of the momentum operator satisfy
\begin{equation}
\begin{gathered}
  P_\nu  \left| {p_\nu  } \right\rangle  = p_\nu  \left| p \right\rangle  \\ 
  \left\langle p \right|\left. {p'} \right\rangle  = \delta ^4 \left( {p - p'} \right) \\ 
  \left\langle x \right|\left. p \right\rangle  = \frac{{e^{ - ipx} }}
{{\left( {2\pi } \right)^2 }} \\ 
\end{gathered} 
\end{equation}
It should be noted that $X^{\mu}$ represents an abstract point in 
4-dimensional Minkowski space, and $P_{\nu}$ is the translation 
operator for such points, They do not refer to the position of an 
actual particle.

Let
\begin{equation}
K_0  = \frac{{2\sin ^2 \left( {\pi d_u } \right)}}
{{\pi A_{du} }}
\end{equation}
The action can now be expressed as
\begin{equation}
I = \left\langle {\Phi _u \left| K \right|\Phi _u } \right\rangle 
\end{equation}
where the operator K is
\begin{equation}
K = K_0 \int_0^\infty  {dM^2 \frac{{\left( {M^2 } \right)^{2 - d_u } }}
{{M^2  - P^2  - i\varepsilon }}} 
\end{equation}
The gauge transformation can now be expressed as a transformation of 
vectors and operators:
\begin{equation}
\begin{gathered}
  \left| {\Phi _u } \right\rangle  \to e^{ig\Lambda } \left| {\Phi _u } \right\rangle  \hfill \\
  A_\mu   \to A_\mu   - i\left[ {P_\mu  ,\Lambda } \right] \hfill \\ 
\end{gathered} 
\end{equation}
Here we have
\begin{equation}
\begin{gathered}
  \Lambda  = \int {d^4 x\left| x \right\rangle \Lambda \left( x \right)\left\langle x \right|}  \\ 
  A_\mu   = \int {d^4 x\left| x \right\rangle A_\mu  \left( x \right)\left\langle x \right|}  \\ 
   - i\left[ {P_\mu  ,\Lambda } \right] = \int {d^4 x\left| x \right\rangle \partial _\mu  \Lambda \left( x \right)\left\langle x \right|}  \\ 
\end{gathered} 
\end{equation}
The combination
\begin{equation}
D_\mu   = P_\mu   - gA_\mu  
\end{equation}
is then invariant under the gauge transformation:
\begin{equation}
D_\mu   = e^{ - ig\Lambda } D_\mu  e^{ + ig\Lambda } 
\end{equation}
Replacing $P_{\mu}$ in K by $D_{\mu}$ then gives us a gauge invariant 
action. 

Let
\begin{equation}
G_M \left( {p^2 } \right) = \frac{1}
{{M^2  - P^2  - i\varepsilon }}
\end{equation}

To first order in g, the action is
\begin{equation}
I = I_0  - K_0 g\int_0^\infty  {dM^2 \left( {M^2 } \right)^{2 - d_u } } \left\langle {\Phi _u } \right|\frac{1}
{{M^2  - P^2  - i\varepsilon }}\left\{ {A_\mu  ,P^\mu  } \right\}\frac{1}
{{M^2  - P^2  - i\varepsilon }}\left| {\Phi _u } \right\rangle 
\end{equation}
It is most convenient to evaluate this in momentum space.  With
\begin{equation}
\begin{gathered}
  A_\mu   = \int {d^4 kd^4 ld^4 x\left| k \right\rangle } \left\langle {k}
 \mathrel{\left | {\vphantom {k x}}
 \right. \kern-\nulldelimiterspace}
 {x} \right\rangle A_\mu  \left( x \right)\left\langle {x}
 \mathrel{\left | {\vphantom {x l}}
 \right. \kern-\nulldelimiterspace}
 {l} \right\rangle \left\langle l \right| \\ 
   = \int {d^4 ld^4 q\left| {l + q} \right\rangle } A_\mu  \left( q \right)\left\langle l \right| \\ 
\end{gathered} 
\end{equation}

where
\begin{equation}
A_\mu  \left( q \right) = \int {\frac{{d^4 x}}
{{\left( {2\pi } \right)^4 }}e^{iqx} A_\mu  \left( x \right)} 
\end{equation}
and with also
\begin{equation}
\Phi _u \left( p \right) = \int {\frac{{d^4 x}}
{{\left( {2\pi } \right)^4 }}\Phi _u \left( x \right)} 
\end{equation}
so that
\begin{equation}
\left| {\Phi _u } \right\rangle  = \left( {2\pi } \right)^2 \int {d^4 p\left| p \right\rangle } \Phi _u \left( p \right)
\end{equation}
then
\begin{equation}
I_1  =  - K_0 g\int_0^\infty  {dM^2 \left( {M^2 } \right)^{2 - d_u } \int {d^4 p'd^4 qd^4 p\left( {2\pi } \right)^4 \delta \left( {p' - p - q} \right)} } \Phi _u^ *  \left( {p'} \right)D_M \left( {p'} \right)A_\mu  \left( q \right)\left( {p'^\mu   + p^\mu  } \right)D_M \left( p \right)\Phi _u \left( p \right)
\end{equation}
Defining the vertex function as
\begin{equation}
ig\Gamma ^\mu  \left( {p',q,p} \right)\left( {2\pi } \right)^4\delta ^4 \left( {p' - q - p} \right) = \frac{{i\delta ^3 I}}
{{\delta \Phi _u^\dag  \left( {p'} \right)\delta A_\mu  \left( q \right)\delta \Phi _u \left( p \right)}}
\end{equation}
We get
\begin{equation}
\begin{gathered}
  i\Gamma ^\mu   =  - i\frac{{2\sin \left( {\pi d_u } \right)}}
{{A_{du} }}e^{i\pi d_u } \frac{{p'^\mu   + p^\mu  }}
{{p'^2  - p^2 }}\left( {\left( {p'^2 } \right)^{2 - d_u }  - \left( {p^2 } \right)^{2 - d_u } } \right) \\ 
   = \frac{{2p^\mu   + q^\mu  }}
{{2p \cdot q + q^2 }}\left[ {S^{ - 1} \left( {p'} \right) - S^{ - 1} \left( p \right)} \right] \\ 
\end{gathered} 
\end{equation}
The same as Terning et al's result.~\cite{Tern}

\section{Conclusions}\label{S:conc}
We have found an operator method that can be used to introduce gauge fields into 
the unparticle effective action.  This method yields results for the 
vertex that are considerably simpler than those obtained by rigorous 
application of the path integral method and are the same as those 
obtained from the looser path integral method of Terning et 
al.~\cite{Tern} We would like to point out that there are however 
ordering problems with the operator method.  An action of the form
\begin{equation}
I  = \int {d^4 xd^4 y\partial _{x\mu } \Phi _u^\dag  \left( x \right)} K\left( {x,y} \right)\partial _y^\mu  \Phi _u \left( y \right)
\end{equation}
where K here involves one less power of $P^{2}$, can be shown to 
yield a quite different vertex.

\section{Acknowledgements}\label{S:ack}
I would like to express my thanks to Wai-Yee Keung for interesting me 
in this problem.


\begin{thebibliography}{9}
    \bibitem{Tern}
    Giacomo Cacciapaglia, Guido Marandella and John Terning,
    \emph{Colored Unparticles},
    arXiv:0708.005 [hep-ph].
    \bibitem{Geo-1}
    Howard Georgi,
    \emph{Unparticle Physics},
    arXiv:hep-ph/0703260.
    \bibitem{Geo-2}
    Howard Georgi,
    \emph{Another Odd Thing About Unparticle Physics}
    arXiv:0704.2457 [hep-ph].
    \bibitem{Li}
    A. Lewis Licht,
    \emph{Gauge Fields and Unparticles}
    arXiv:0801.0892 [hep-th]
\end{thebibliography}
\end{document}